\documentclass[twocolumn,amsmath,amssymb,superscriptaddress,pra]{revtex4-1}

\usepackage{graphicx}
\usepackage{dcolumn}
\usepackage{bm}
\usepackage{epsf}
\usepackage{amsmath,txfonts}
\usepackage{amssymb}
\usepackage{dsfont}
\usepackage{color}
\usepackage[colorlinks=true,linkcolor=blue,urlcolor=blue,citecolor=blue,pdfusetitle]{hyperref}
\usepackage{hyperref}

\newcommand{\bra}[1]{\langle #1|}
\newcommand{\ket}[1]{|#1\rangle}

\newcommand{\tr}[1]{\mathrm{tr}\left[#1\right]}
\newcommand{\ptr}[2]{\mathrm{tr_{#1}}\left[#2\right]}

\newcommand{\bla}{bla\\bla\\bla\\bla}
\newcommand{\more}{more\\more\\more\\more\\more}

\newcommand{\PRA}{Phys. Rev. A }

\newcommand{\PRE}{Phys. Rev. E }

\newcommand{\PRL}{Phys. Rev. Lett. }

\newcommand{\NJP}{New. J. Phys. }

\begin{document}

\title{Implications of non-Markovian dynamics on information-driven engine}

\author{Obinna Abah}\email[]{o.abah@qub.ac.uk}  \author{Mauro Paternostro}\email[]{m.paternostro@qub.ac.uk}
\affiliation{Centre for Theoretical Atomic, Molecular and Optical Physics, School of Mathematics and Physics, Queen's University Belfast, Belfast BT7 1NN, United Kingdom}

\begin{abstract}
The understanding of memory effects arising from the interaction between system and environment is a key for engineering quantum thermodynamic devices beyond the standard Markovian limit. We study the performance of measurement-based thermal machine whose working medium dynamics is subject to backflow of information from the reservoir via collision based model. In this study, the non-Markovian effect is introduced by allowing for additional unitary interactions between the environments. We present two strategies of realizing non-Markovian dynamics and study their influence on the performance of the engine. Moreover, the role of system-environment memory effects on the engine work extraction and information gain through measurement can be beneficial in short time.
\end{abstract}

\maketitle

\section{Introduction}

The second law of thermodynamics is ubiquitous in nature: it stipulates that heat always flows from hot place to cold one. However, in 1867 Maxwell proposes the opposite with his idea of an intelligent demon to illustrate the statistical nature of the second law of thermodynamics \cite{lef03}.
The demon, with sufficiently information about the microscopic motions of individual atoms and molecules, is capable to separate the fast-moving (``hot") ones from the slow-moving (``cold") ones and induce the heat to flow from cold to hot, in apparent contradiction with the second law of
thermodynamics.
It took nearly a century to resolve this apparent paradox following a series of works, starting from Szilard's engine \cite{szi29} through Landauer \cite{lan61}, Bennett \cite{ben82} and others to clarify the link between the information recorded by the demon and the thermodynamic entropy, see \cite{par15}. 
The advances in nanotechnology have made the realization of Maxwell's thought experiment, Szilard's engine possible in recent time \cite{toy10,ber12,kos14,kos14a}. 

In addition to this, there has been a parallel line of development in the non-Markovian dynamic behavior of system interacting with reservoir. Theoretical advances have been made on its characterization \cite{bre09,riv10,lor13}  as well as verifications  \cite{liu11,chi12,sou13} in various experimental setup. The role of memory (non-Markovian) effects in understanding of information processing at both the classical and quantum level is currently attracting research interest \cite{ree14,lor15,pez16,ham17}.  Likewise, over the last few years , there has been an increase on the studies to understand or harness the non-Markovian effect on quantum thermodynamic machines \cite{gel13,che16,tho18}. Recently, studying the non-Markovian dynamic of a system has shed more light into the understanding of the Landauer principle \cite{pez16}.

Over the past few years, great effort has been devoted on  studying the interplay between thermodynamics and quantum mechanics \cite{jem04,hub08,cam11,maz13,bat14,an15}. Remarkable progress has been made in understanding the non-equilibrium processes in thermodynamics \cite{jar11} as well as extending/generalizing the second law of thermodynamics to incorporates measurement and feedback driven processes \cite{sag08,jac09,sag10,def13,fun13,hor14,goo16,str17}. Recently, the role of feedback control on information thermodynamic engine has been experimentally studied in different platform \cite{cam16,cia17,cot17,mas17,xio18,nag18}. However, the understanding of the machine performance when the feedback engine protocol is performed by  system exhibiting non-Markovian dynamics  is still lacking.  Although the self-consistent formulation of an interpretation of thermodynamic laws in the presence of measurements and feedback is still work in progress, and is attracting much attention, more practical issues such as the enhancement of the performance of cooling algorithms by feedback-based mechanisms are already under investigation and exploitation~\cite{boy02,bau05,liu16,rod17}.

In this paper, we investigate the implications of  non-Markovian dynamics on feedback information-driven machines. Our findings show that memory effect can enhance the overall performance --- work extraction and information gain of the engine in a short time. The rest of the paper is organized as follows. In the next section, Section \ref{section2}, we first present the description of the measurement-based engine and then briefly discuss its thermodynamic analysis. In Section \ref{section3} we  introduce the collision based model of realizing non-Markovian dynamics and outline example of two different strategies. Then, the characterization of the non-Markovian features is numerical analyze in Section \ref{section4a}, while   the analysis of the feedback-driven engine in both Markovian and non-Markovian situation is devoted to Section \ref{section4b}. Finally, Section \ref{section5} draws our conclusions.
\begin{figure}[!]
 \includegraphics[width=\columnwidth]{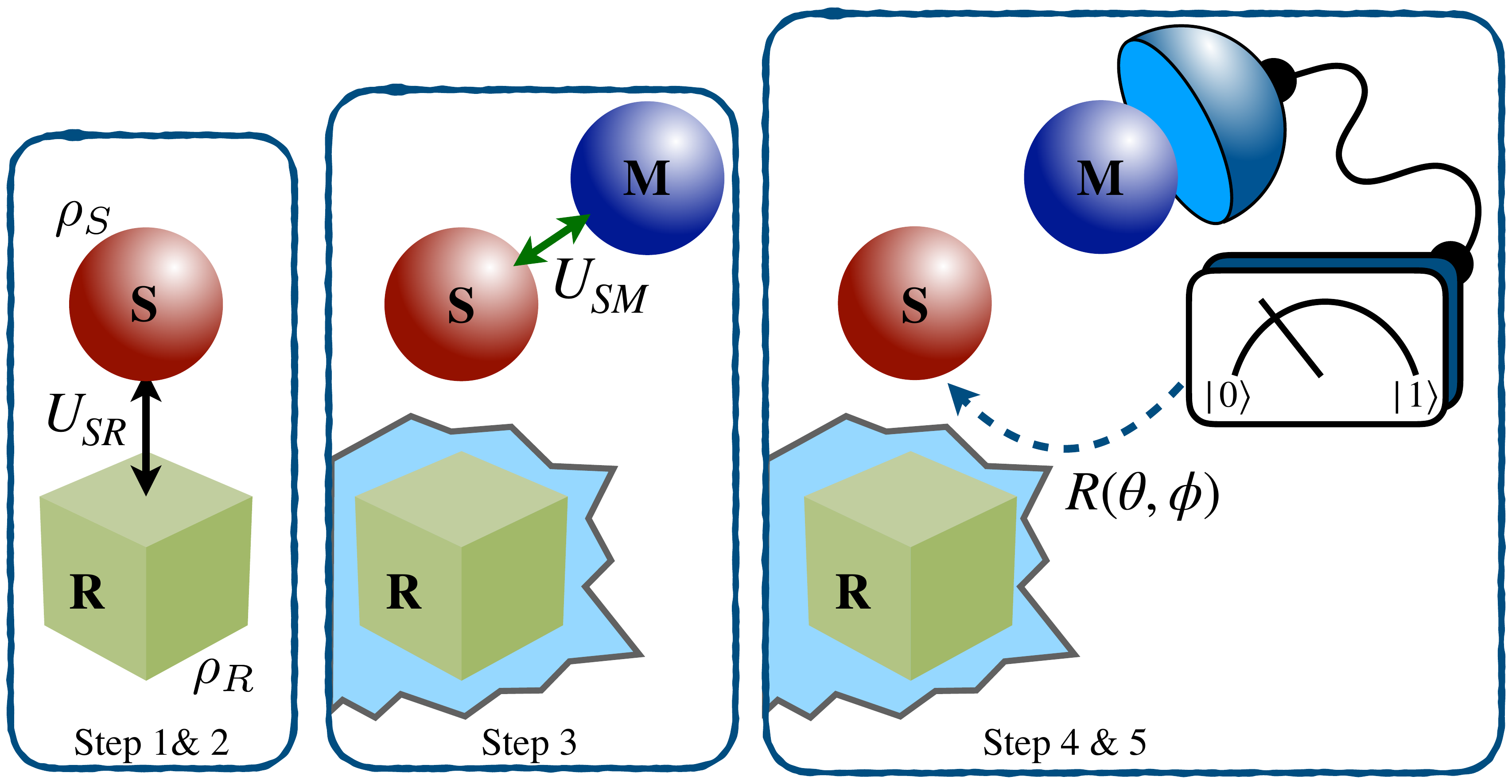}
 \caption{The various steps of the general protocol that we consider. The jagged light-blue area stands for the tracing-out of the environmental system. The information-gathering process in Step 3 consists of a projective measurement performed on the state of the ancilla $M$, which is projected onto the elements of its computational basis, such as $\{\ket{0}_M,\ket{1}_M\}$ in the case of a qubit.}
 \label{protocol}
\end{figure}

\section{Measurement-based thermo-machine}
\label{section2}
The system initially prepared in given state is brought to contact with a heat reservoir. Then, the system is decoupled and attached to a measuring apparatus initially prepared in a given state. The apparatus determine the state of the system and depending on the result of the measurement, a feedback operation is performed on the system. The setup consists of three components; system,  reservoir and an ancilla.

\subsection{Description of the protocol}
\label{section2a}
We now introduce and illustrate the protocol that we aim at studying for the investigation of the effects that a process of information-gathering and feedback have on the capability of the system to perform work. We proceed step by step, as follows:

\noindent
\textit{Step 1: Initial preparation.--}
System $S$ and thermal reservoir(s) $R$ are prepared in their respective equilibrium states at inverse temperature $\beta_i = 1/k_BT_i$ and frequency $\omega_i$, where $i = S,R$.  The initial system-reservoir state is described by the density matrix
\begin{equation}
\rho_{SR} = \bigotimes_{i=R,S}\rho_i =\bigotimes_{i=R,S} \frac{e^{-\beta_i H_i}}{Z_i},
\end{equation}
where $H_{i}$ denotes the Hamiltonian of element $i$ and $Z_i = \tr{e^{-\beta_i H_i}}$ is the corresponding partition function.
For simplicity, we will consider the case in which the system and the reservoir are made of two-level systems. 

\noindent
\textit{Step 2: System-environment coupling.--}
System and reservoir interact unitarily. In line with the usual formalism used in collisional models for quantum open-system dynamics~\cite{str17,sca02,zim02,ciccarello,lorenzo,strunz,sea18}, in what follows we will concentrate on a time-evolution operator of the partial-SWAP form such as
\begin{equation}
\label{USR}
U_{SR} = e^{-i \tau}\left[ \cos(2 \tau) \mathds{1}_4 + i \sin (2 \tau) U_{sw} \right],
\end{equation}
where $\tau$ is a dimensionless interaction time and $U_{sw}$ is the two-particle SWAP transformation $\ket{i,j}_{SR}\stackrel{U_{sw}}{\longrightarrow}\ket{j,i}_{SR}$ with $\ket{i}_S$ [$\ket{j}_R$] a state of the computational basis chosen for $S$ [$R$].
The $S$-$R$ state after such unitary evolution is thus 
\begin{equation}
\rho^{u}_{SR} = U_{SR} (\rho_{S} \otimes \rho_R) U_{SR}^\dagger.
\end{equation}
 In general, the joint dynamics embodied by $U_{SR}$ gives rise to quantum correlations between system and environment. The environment is then discarded, leaving us with the reduced state of the system only
\begin{equation}
\rho^u_{S} = \ptr{R}{\rho^u_{SR}}.
\end{equation}
\noindent
\textit{Step 3: Pre-measurement.--} 
The system is then brought into contact with a measuring apparatus, i.e. an ancillary qubit $M$ prepared in state $\rho_M$. The $S$-$M$ coupling takes place according to the unitary transformation $U_{SM}$, which gives the joint density matrix
\begin{equation}
\rho^{pm}_{SM} = U_{SM} (\rho^u_{S} \otimes \rho_M) U_{SM}^\dagger.
\end{equation}
We assume that $U_{SM}$ takes place over a dimensionless system-probe interaction time $\tau_m$ and that the corresponding coupling Hamiltonian $H_{SM}$ such that $U_{SM}\!=\!e^{-i\tau_m (H_{SM} + H_S)}$ takes the form of a spin-spin coupling $ H_{SM}\!=\!\sigma^{(j)}_S \otimes \sigma^{(j)}_M$, whose form will be specified later on. Here, $\sigma^{(j)}_i$ is the $j=x,y,z$ Pauli spin operator of particle $i\!=\!S,M$.

\noindent
\textit{Step 4: Measurement.--} This is the actual information-gathering step where the information on $S$ acquired by the ancilla during Step 3 through their mutual interaction is inferred via an actual measurement process. The latter is described by the complete set of projective operators $\left\{M^{(k)}_M\right\}$, defined in the Hilbert space of the ancilla $M$. Let us assume that the ancilla is initially prepared in one of its computational-basis states, i.e. $\rho^M_p=\ket{p}\bra{p}_M$. The probability that outcome $k$ is obtained as a result of such measurement  is given by
\begin{equation}
P_k = \ptr{SM}{M^{(k)}_M \rho^{pm}_{SM} M^{(k)}_M}=\ptr{S}{{\cal F}_k\rho^u_S}
\end{equation}
with ${\cal F}_k={\cal E}^\dag_k{\cal E}_k$ and ${\cal E}_k=\!{}_M\bra{k}U_{SM}\ket{p}_M$ an element of the positive-operator value measure (POVM) induced on the system. The corresponding post-measurement state of the system reads
\begin{equation}
\rho^{k}_S=\frac{{\cal E}_k\rho^u_S{\cal E}^\dag_k}{P_k}.
\end{equation}
\noindent
\textit{Step 5: Feedback control operation.--} Based on the outcome of the measurement at Step 4, the controller performs a conditional operation on the state of the system~\cite{sag08,fun13}. The most general unitary transformation on a single-qubit state is a rotation by an angle $\alpha$ about an arbitrary axis identified by the unit vector ${\bf n}=(\sin\theta\cos\phi,\sin\theta\sin\phi,\cos\theta)$, which has been written in polar coordinates specified by the polar angle $\theta$ and azimuthal one $\phi$. By including a general global phase $\gamma$, such rotation reads
\begin{equation}
R({\bm v})=
e^{i\gamma}\left(
\begin{array}{cc}
 \cos \frac{\alpha }{2}-i \cos\theta\sin\frac{\alpha }{2}
   &-i \sin \frac{\alpha }{2} \sin \theta  e^{-i\phi} \\
-i \sin \frac{\alpha }{2} \sin \theta  e^{i\phi}& \cos
   \frac{\alpha }{2}+i \cos \theta  \sin \frac{\alpha }{2} \\
\end{array}
\right)
\end{equation}
with ${\bm v}:=(\gamma,\alpha,\theta,\phi)$. In our case, the set of parameters upon which such rotation depends should be interpreted as conditioned on the outcome of the measurement performed, at Step 4, on the ancilla $M$. That is
\begin{equation}
{\bm v}\longrightarrow{\bm v}_k:=(\gamma_k,\alpha_k,\theta_k,\phi_k).
\end{equation}
The use of such conditioned rotation, which embodies our simple feedback control operation, delivers the state of the system 
\begin{equation}
\rho^{fb}_{S,k} = R({\bm v}_k) \rho^k_S R^\dagger({\bm v}_k).
\end{equation}

\noindent
\textit{Step 6: The reset.--} The system evolves independently and a fresh ancilla is made available to the next iteration of the protocol, which proceeds again from Step 1 onwards. This stage has no effect on the analysis that follows.

\subsection{Thermodynamics of the machine}
\label{section2b}
We proceed with the thermodynamic analysis of the protocol presented above, by calculating the changes in internal energy $E[\rho] \equiv \tr{H\rho}$ and entropy $S[\rho] \equiv - k_B \tr{\rho \ln \rho}$ of the system associated with the preparation, measurement and feedback-control protocols. 

First, after the system preparation (interaction with the reservoir), the change in the system internal energy is 
\begin{equation}
\Delta E^{u} = E[\rho^u_{S}] - E[\rho_{S}] = \tr{H_S\,\rho^u_{S}} - \tr{H_S\, \rho_{S}},
\end{equation}
and the change in system entropy reads
\begin{equation}
\Delta S^{u}\!=\!S[\rho^u_{S}] - S[\rho_{S}]\!=\!-k_B\left(\tr{\rho^u_{S} \ln\rho^u_{S}} - \tr{\rho_{S} \ln\rho_{S}}\right).
\end{equation}
From the first law of thermodynamics,  $\Delta E\!=\!\Delta W + \Delta Q$, and assuming that the heat exchange between the system and reservoir is governed by $\Delta Q^u_S\!=\!-\Delta Q^u_R$, the work done on/by the system can be written as $\Delta W^{u}\!=\!\Delta E^{u} + \Delta Q^{u}_R$, where the last term associates the exchange of heat $\Delta Q^u_R\!=\!\tr{H_R \left(\rho_R^u - \rho_R\right)}$ with its environment and $\rho_R^u\!=\!\ptr{S}{\rho^u_{SR}}$ is the marginal state of the reservoir after interaction.

For the measurement step, the thermodynamic quantities are as follows. The variation of internal energy of the system reads
\begin{equation}
\Delta E^{pm}\!=\!E[\rho^{pm}_{S}] - E[\rho^u_{S}]\!=\!\tr{H_S \rho^{pm}_{S}} - \tr{H_S \rho^u_{S}},
\end{equation}
where $\rho^{pm}_S\!=\!\ptr{M}{\rho^{pm}_{SM}}$ is the reduced state of the system after the pre-measurement step (cf. Step 3).
The corresponding change in entropy of the state of the system is
\begin{equation}
\Delta S^{pm}\!=\!S[\rho^{pm}_{S}] - S[\rho^u_{S}]\!=\!-{k_B}\left(\tr{\rho^{pm}_{S} \ln\rho^{pm}_{S}} - \tr{\rho^u_{S} \ln\rho^u_{S}}\right).
\end{equation}
Based on the second law of phenomenological non-equilibrium thermodynamics, the entropy production characterizing the irreversibility of the measurement process reads~\cite{str17}
\begin{equation}
\label{sigma}
\begin{aligned}
 \Sigma^{m} &\!=\!\Delta S_{SM}^{m} = \Delta S_S^{m} +\Delta S_M^{m} - I_{S:M}^{m} =0.
\end{aligned}
\end{equation}
The first two terms in the right-hand side Eq.~\eqref{sigma} correspond to the change in entropy of the system and the ancilla, while the third term is the quantum mutual information between them.  As the state of the measurement apparatus is unchange, we have $\Delta S_M^{m}\!=\!0$, in this particular case. The gain of information about the system achieved through the measurement is $I_{S:M}^{m} = S(\rho_S^k) - S(\rho_S^u) \equiv I_{qm}$. On the other hand, from the first law, we have that the work done by the measurement reads $\Delta W^{m}\!=\!\Delta E^{m} \!=\! E[\rho_S^k] - E[\rho_S^{u}]$.

Then, during the feedback step, the variation of system energy and entropy are
\begin{equation}
\begin{aligned}
\Delta E^{fb}_k &\!=\! E[\rho^{fb}_{S,k}] - E[\rho^{k}_{S}]\!=\!\tr{H_S \rho^{fb}_{S,k}} - \tr{H_S \rho^{k}_{S}},\\
\Delta S^{fb}_k &\!=\!  S[\rho^{fb}_{S,k}] - S[\rho^{k}_{S}],
\end{aligned}
\end{equation}
respectively. 
Here definitions analogous to those used above hold for both $S[\rho^{fb}_{S,k}]$ and $S[\rho^{u}_{S}]$. Using again the first and second laws, we have
\begin{equation}
\Delta E^{fb}_k\!=\!\Delta W^{fb}_k + \Delta Q^{fb}_k,~~\Sigma^{fb}\!=\! \Delta S_k^{fb}  +  \Delta S_c^{fb} \ge 0,
\end{equation}
where   $\Delta S^{fb}_k $ and $\Delta S_c^{fb}$ are the entropy change associated with  the system and  feedback controller respectively. Thus, an upper bound to the amount of thermodynamic work extracted by the feedback protocol is  $ \Delta W^{fb}_k \ge\Delta E^{fb}_k - T_c \Delta S_k^{fb}$, where we employ the relation $\Delta S\!=\!-\Delta Q/T_c$ and $T_c$ is the controller temperature.
The work extracted by the system is beyond the second law due to the correlation between the system and the memory. The form of such bound was first given in Ref.~\cite{sag08}, more details on the subject can be found in Ref.~\cite{str17}. We note that the feedback protocol can be engineered so as  to change only the system density matrix and leave that of the ancilla unaffected. This is possible, for instance, by post-selecting only the measurement events where the state of the ancilla is found to be the initially prepared one $\ket{p}_M$.  Therefore,  neglecting $-\Delta S^{fb}_k$ which is usually non-negative, we defined the total work done on/by the system through the measurement and feedback protocol as
 \begin{equation}
  \Delta W_{t} =  \Delta W^m + \Delta W^{fb}  = E[\rho^{fb}_{S,k}] - E[\rho^{u}_{S}].
 \end{equation}


\section{Non-Markovian dynamics of the system - collisional based model}
\label{section3}
Here, we consider a situation where the system undergoes non-Markovian dynamics as a result of its interaction with the environment (taking place at steps 1 and 2 of our protocol). The realization of the dynamics that we decide to consider is that of collisional models, which offer great flexibility and richness of phenomenology~\cite{ciccarello}. 

In particular, we consider the case in which the reservoir's memory mechanism arises from collisions between different elements of a structured, multi-party environment, following an interaction with the system. This scenario has been successfully used in the past to model memory-bearing mechanisms able to propagate to the environment information acquired on the state of the system~\cite{mcc14}. More recently, this realization of memory-bearing effects has been used to assess the performance of a quantum Otto cycle having a harmonic system as a working medium~\cite{pezzuttoQST}. Collisional models allow for the tracking of the dynamics of both system and environments, which in turn makes it possible to follow the ensuing emergence of the system-environment correlations responsible for memory effects~\cite{sca02,ciccarello,lorenzo,strunz,cam18,cakmakPRA2019}. They are thus invaluable methodological tools to assess the back-action of memory-bearing environments on the information-driven engine at the core of our study. 

\begin{figure}[!t]
\includegraphics[width=\columnwidth,angle=0]{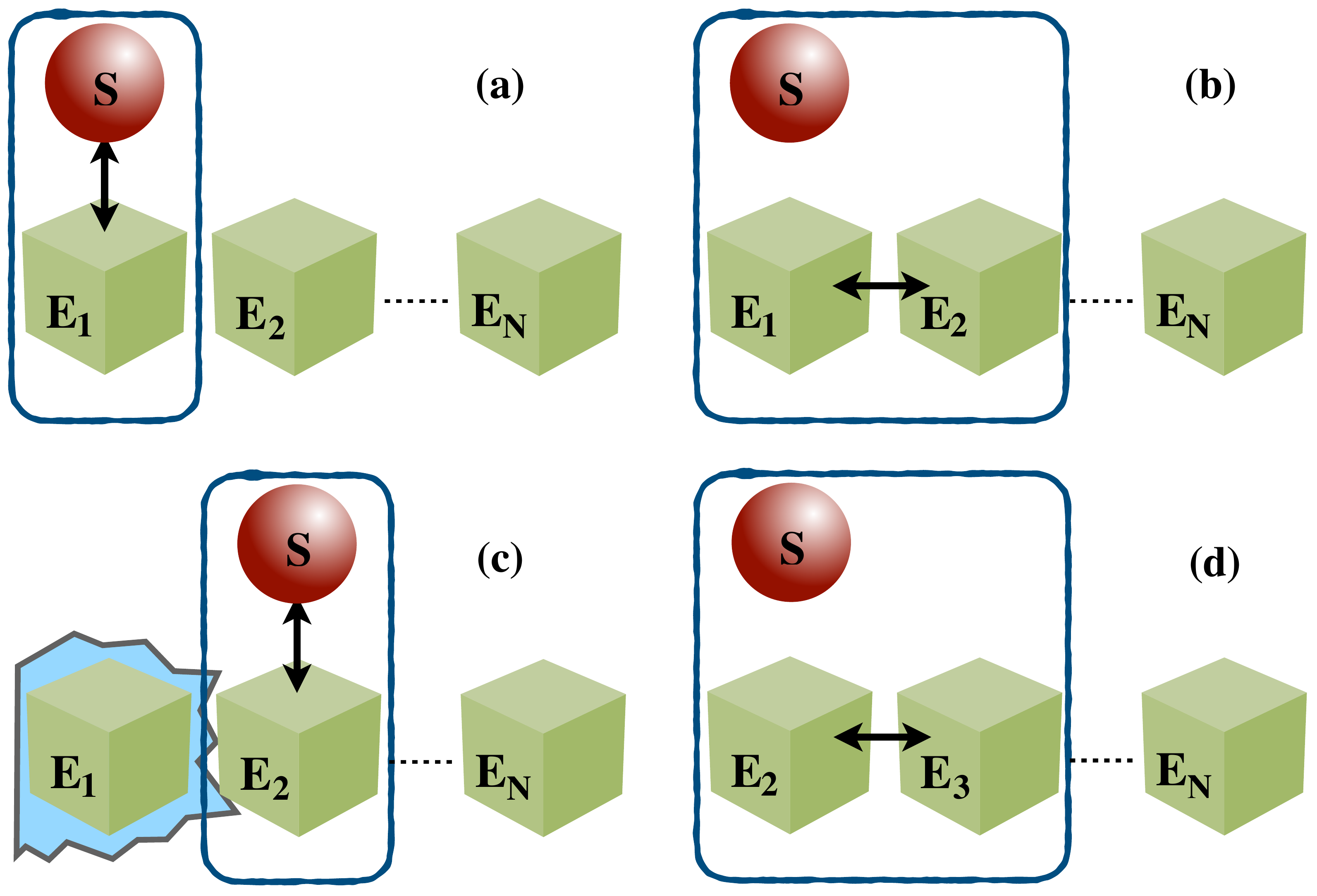}
\caption{Schematic of non-Markovian dynamics via collision model for nearest sub-environment collisions. The system and the sub-environment particles are initially uncorrelated. In the first step (a), the system $S$ interacts with $E_1$. The next step, (b) $E_1$ interacts with $E_2$ and thereby correlating the system and particles $E_1$ and $E_2$. Then step (c), $E_1$ is traced away. After which  the system interacts with $E_2$ before isolating the system for measurement and feedback processes in strategy 1. For the strategy 2, the system and sub-environment particles collisional iterations are performed up to $E_3$, (a) - (d), before the measurement and feedback.}
\label{collision model}
\end{figure}

As anticipated above, we assume an environment $R$ made up of a large number of elements, which we label $\{E_1, E_2,..,E_n\}$ and assume, for the sake of simplicity, to be mutually identical. The total state of system and environment is initially factorized and the dynamics proceeds through as sequential collisions (interaction process) between $S$ and an element $E_n$ of the environment. These are followed by pairwise collisions/interactions between the elements of the-environment, as illustrated in Fig.~\ref{collision model}. In Ref.~\cite{mcc14}, it has been shown that the degree of non-Markovianity of the reduced system dynamics depends on how the erasure of system-environment correlations is performed. 

Here, we will consider two inequivalent schemes of tracing out the degree of freedom of the environment. The first scenario that we consider to compute the reduced dynamics of $S$ requires the  environmental particle $E_n$ to be traced out when it has interacted with $S$ and $E_{n+1}$ but before the system interacts with $E_{n+1}$. In the second scenario,  the reduced dynamics of the system is obtained by tracing out the environmental particle once it has interacted with system $S$. The remaining environmental particle interacts with the next homonimous particle before the latter subsequent collides with the system.
We also assume that the environment-environment interaction evolution is described by the unitary operator~\cite{ciccarello,lorenzo,strunz}
\begin{equation}
\label{UEE}
U_{EE} = e^{-i \tau_e}\left[ \cos(2 \tau_e) \mathds{1}_4 + i \sin (2 \tau_e) U_{sw} \right],
\end{equation}
which describes another partial-SWAP gate between two consecutive elements of the environment,  parameterized by the dimensionless interaction time  $\tau_e$ .

The first scenario (which we term \textit{strategy-1}) that we consider involves tracing out the particle $E_n$ after it has collided with $E_{n+1}$, as exemplified in Fig.~\ref{collision model} {\bf (a)} - {\bf (c)}. It starts with a collision between $S$ and $E_n$, modelled through the unitary operation $U_{SR}$ in Eq.~(\ref{USR}), which delivers the joint state 
\begin{equation}
\rho_{SE_n} = U_{SR} (\rho_S \otimes \rho_{E_n}) U_{SR}^\dagger.
\end{equation}
The three particles $S, E_n$ and $E_{n+1}$ then become correlated through the intra-environment interaction $U_{E_n E_{n+1}}$ in Eq.~\eqref{UEE}, after which particle $E_n$ is traced out. This results in the bipartite $S$-$E_{n+1}$ state 
\begin{equation}
\rho_{SE_{n+1}} = {\rm tr}_{E_n}\left[U_{EE}(\rho_{SE_n} \otimes \rho_{E_{n+1}}) U_{EE}^\dagger\right].
\end{equation}
The marginal state of the system is computed after the interaction with $E_{n+1}$. Thus, \textit{strategy-1} prepare the system in state 
\begin{equation}
\rho_{S}^u = \mathrm{tr}_{E_{n+1}}\left[U_{SR} \, \rho_{SE_{n+1}} U_{SR}^\dagger\right].
\end{equation}
 We remark that retaining the correlations up to the third environment -- which corresponds to the systematic collision with the environmental components $E_n, E_{n+1},$ and $E_{n+2}$ as in Fig.~\ref{collision model} -- does not change the resulting dynamics \cite{cam18}.
At the end of the system-environment interaction, the engine-protocol steps [$\textit{step 3 - 6}$] are performed before the system collides with another fresh environment.

In the second scenario, dubbed \textit{strategy-2}, the correlation established between $S$ and $E_n$ is removed before the intra-environment interaction $E_n - E_{n+1}$. The states achieved at each stage of \textit{strategy-2} are thus as follows. First, the collision between system and $E_n$ occurs, which gives the state
\begin{equation}
\rho_{SE_n} = U_{SR} (\rho_S \otimes \rho_{E_n}) U_{SR}^\dagger, 
\end{equation}
and their resulting marginals $\rho_{S^\prime} = \mathrm{tr_{E_n}} [\rho_{SE_n}]$ and $\rho_{E_{n}^\prime} = \mathrm{tr_{S}} [\rho_{SE_n}]$ for the system and $E_n$ respectively. Then, the marginal state of the $E_{n+1}$ sub-environment component after the intra-environment collision is 
\begin{equation}
\rho_{E^\prime_{n+1}} = \mathrm{tr_{E_n^\prime}} [U_{EE} (\rho_{E_{n}^\prime} \otimes \rho_{E_{n+1}})  U_{EE}^\dagger].
\end{equation}
The resulting state of the system prepared by \textit{strategy-2} becomes
\begin{equation}
\rho_S^u = \mathrm{tr_{E^\prime_{n+1}}} [U_{SR} \, (\rho_{S^\prime} \otimes \rho_{E_{n+1}^\prime}) U_{SR}^\dagger].
\end{equation}
This scenario clearly differs from the first one in both the number of particles being involved, and the amount of correlations that are retained as a result of the system-environment interaction. In turn, this influences the non-Markovian features of the dynamical maps applied to $S$ and arising from the implementation of such strategies. 
 

To quantify the degree of non-Markovianity of the reduced system dynamics undergone by $S$, we employ the measure for non-Markovianity proposed in Ref.~\cite{bre09} which is associated with back-flow of information from the environment to the system. This is based on the time behavior of the trace distance between two different initial quantum states of $S$, that is
\begin{equation}
D(\rho_1,\rho_2)= \frac{1}{2} ||\rho_1 - \rho_2||, 
\label{BLPmeasure}
\end{equation}
where  $||\rho|| = \tr{\sqrt{\rho^\dagger\rho}}$ is the trace norm of operator $\rho$ and $\rho_{1,2}$ are two density matrices of $S$. For Markovian dynamics, $D(\rho_1,\rho_2)$ monotonically decreases with time for any pair of initial states $\rho_{1,2}(0)$. On the contrary, a dynamical process is signalled as  non-Markovian if there is a pair of such states for which this quantity exhibits a non-monotonic behaviour. 
 
 \begin{figure*}[!]
{\bf (a)}\hskip7.5cm{\bf (b)}\\
 \includegraphics[width=0.8\columnwidth]{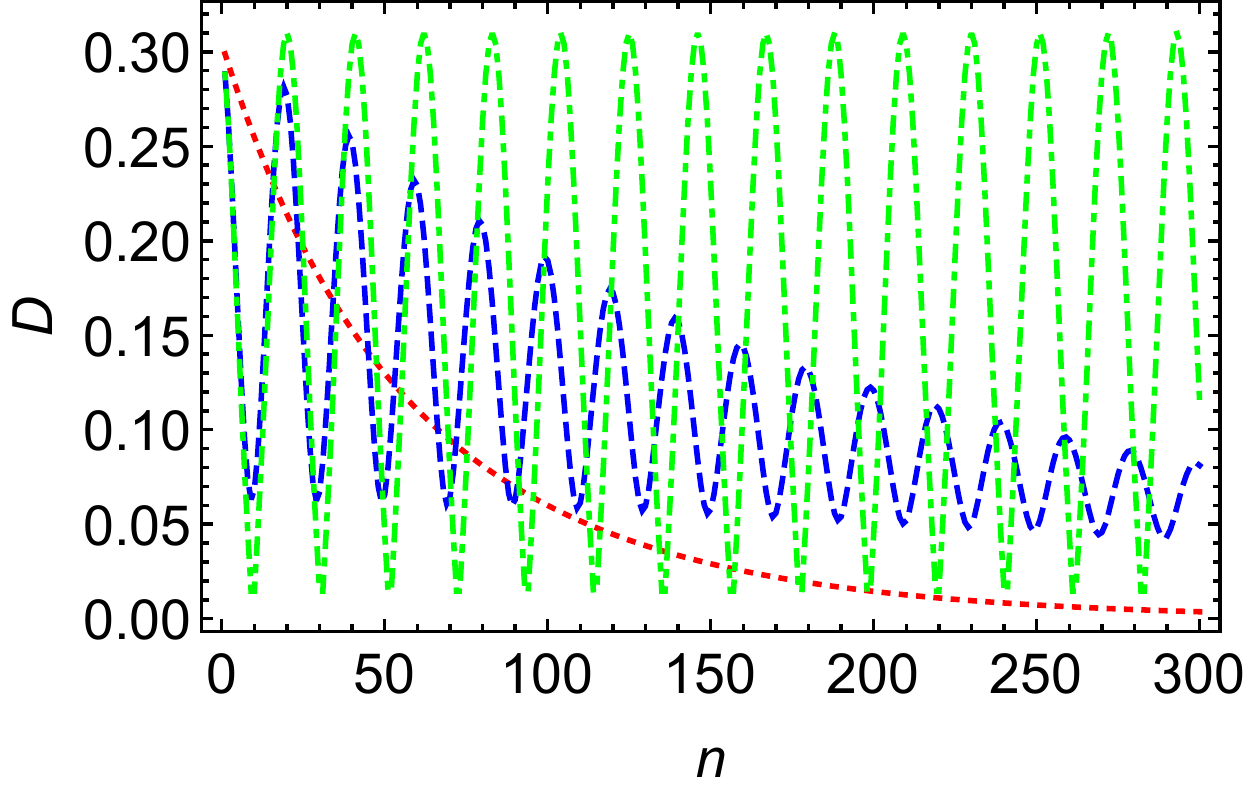}\hfil
 \includegraphics[width=0.8\columnwidth]{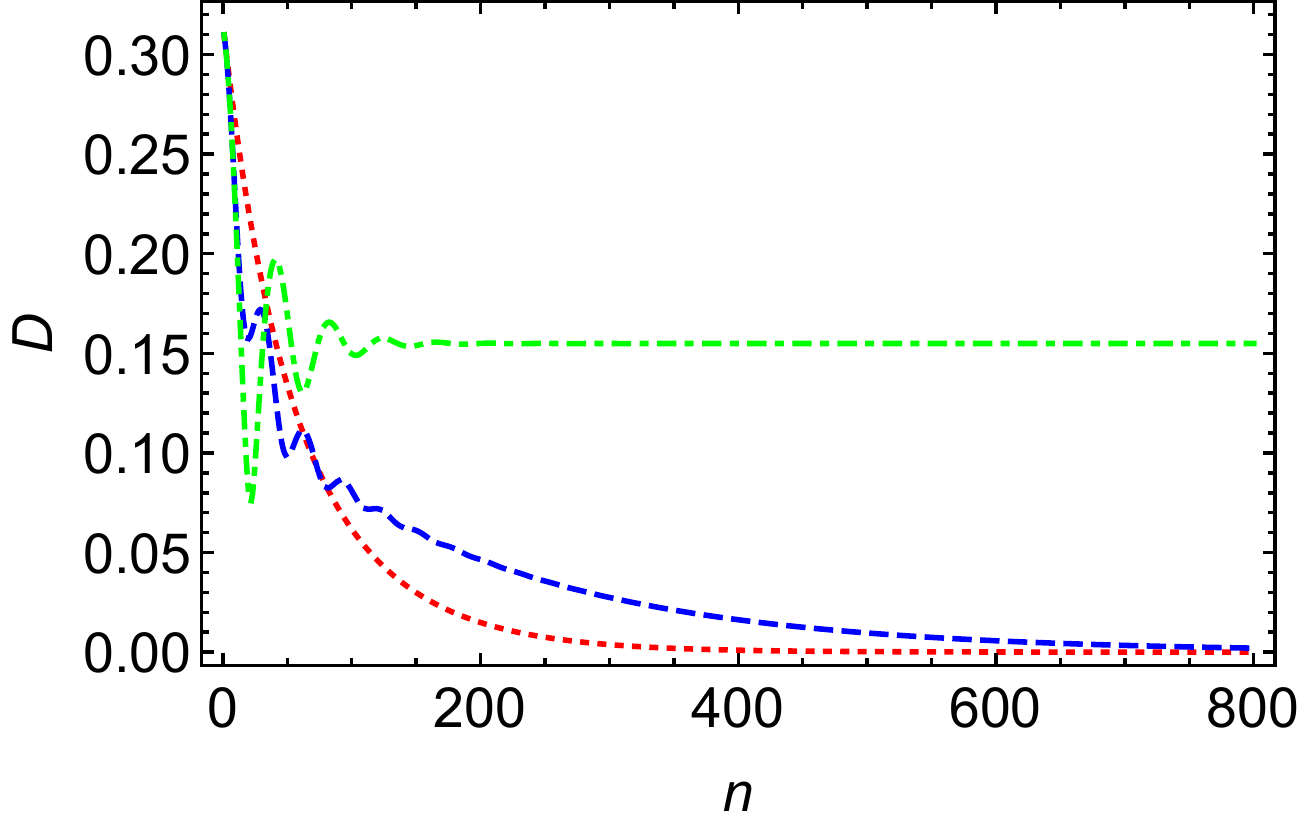}
\caption{The trace distance $D$ between evolved system states as a function of the number of  collision iteration $n$ with the environment for both strategies. Upper  {\bf (a)} [lower {\bf (b)}] panel are the results for the \textit{strategy-1} [\textit{strategy-2}]. We have considered the initial states $\rho_{S1} (\sigma_S^z)$  and $\rho_{S2}(\sigma_S^y)$, while the sub-environments are prepared in $\rho_R(\sigma^z_R)$. The red dotted curve corresponds to the Markovian situation, $\tau_e = 0.0$, while the blue dashed and green dot-dashed curves represent the non-Markovian dynamics with the dimensionless inter-environmental coupling time $\tau_e = 10 \pi/43$ and $\tau_e = \pi/4$ respectively. The system-environment interaction time is $\tau = \pi/42$ for weak coupling and the system and environment frequency parameters are  $\omega_S = 1$ and $\omega_R = 3$, while their inverse temperature is fixed at $\beta_S = \beta_R = 0.94$. }
\label{fig3}
\end{figure*}
\section{Analysis of non-Markovianity and its role in the performance of the engine}\label{section4}
Now we present the numerical analysis of the non-Markovian dynamics of the collision model for both strategies described above and then, their role on the thermodynamics of the engine. In the remainder of the paper, we will assume both the system and reservoir to be   two-level systems with Hamiltonian $H_i= \omega_i\sigma_i^{(j)}/2$ ($i=S,R$), with the thermal state density matrix of the form
\begin{equation}
\rho_{i} \left(\sigma_i^{(j)}\right) = \exp(-\beta_i H_i)/ Z_i, 
\end{equation} 
where $j=x,y,z$ is a label for the $j$-Pauli spin operator of particle $i = S, R$, and $\beta_i$ is the corresponding inverse temperature.

\subsection{Non-Markovianity features from both strategies}
\label{section4a}
We numerically analyze the behaviour of the trace distance $D(\rho_{S_1},\rho_{S_2})$ as the collision-based model for system-environment interactions are repeatedly executed. This analysis elucidates how the system can be initializes in a state resulting in dynamical signatures of non-Markovianity using different strategies described in Section \ref{section3} and corresponds to the first two steps of the engine protocol, see Section \ref{section2a}. We present the behaviour of the trace distance in Eq. (\ref{BLPmeasure}) for two initial states prepared at $\rho_{S1} (\sigma_S^z)$ and $\rho_{S2}(\sigma_S^y)$. We have assumed that all environmental particles/qubits are initialized in the state $\rho_R(\sigma^z_R)$. Figures \ref{fig3}(a) and \ref{fig3}(b) show the differences between the two strategies addressed in this study. For purely Markovian dynamics ($\tau_e = 0$, red dotted curves), the trace distance decreases monotonously while switching on the inter-environment interaction times ($\tau_e \ne 0$, blue dashed and green dot-dashed curves) results in revivals that are evidence of non-Markovianity. In fact, this system-environment interaction produces a backflow mechanism - which is seen as oscillations of the trace distance that fades out in the large number of collisions with fresh ancilla. The strong environment-environment interaction time $\tau_e = \pi/4$ corresponds to a full state-swap between two consecutive environment particles that results in a non vanishing trace distance, see the green dot-dashed curves in Figs. \ref{fig3}(a) and (b). It can be seen that the oscillations are more frequent in \textit{strategy-1} (Fig. 3{\bf (a)}) but fades out to a non-zero value in  the \textit{strategy-2}, see Fig. 3{\bf (b)}. While the non-Markovian dynamics persists for both strategies in strong intra-environment interaction, the intermediate coupling strength shows a clear dependence of the non-Markovian nature on the way information/correlation is developed via collisions. For a weaker environment-environment particle interaction times $\tau_e < \pi/4$, both strategies trace distance decreases as the number of environmental collision increases, see blue dashed curves in Fig. \ref{fig3}. For more extensive discussion on the way information is exchanged between the system and environment for the two strategies and their differences/superiority, see Refs. \cite{mcc14,cam18}.

\begin{figure*}[!]
{\bf (a)} \\
 \includegraphics[width=160pt]{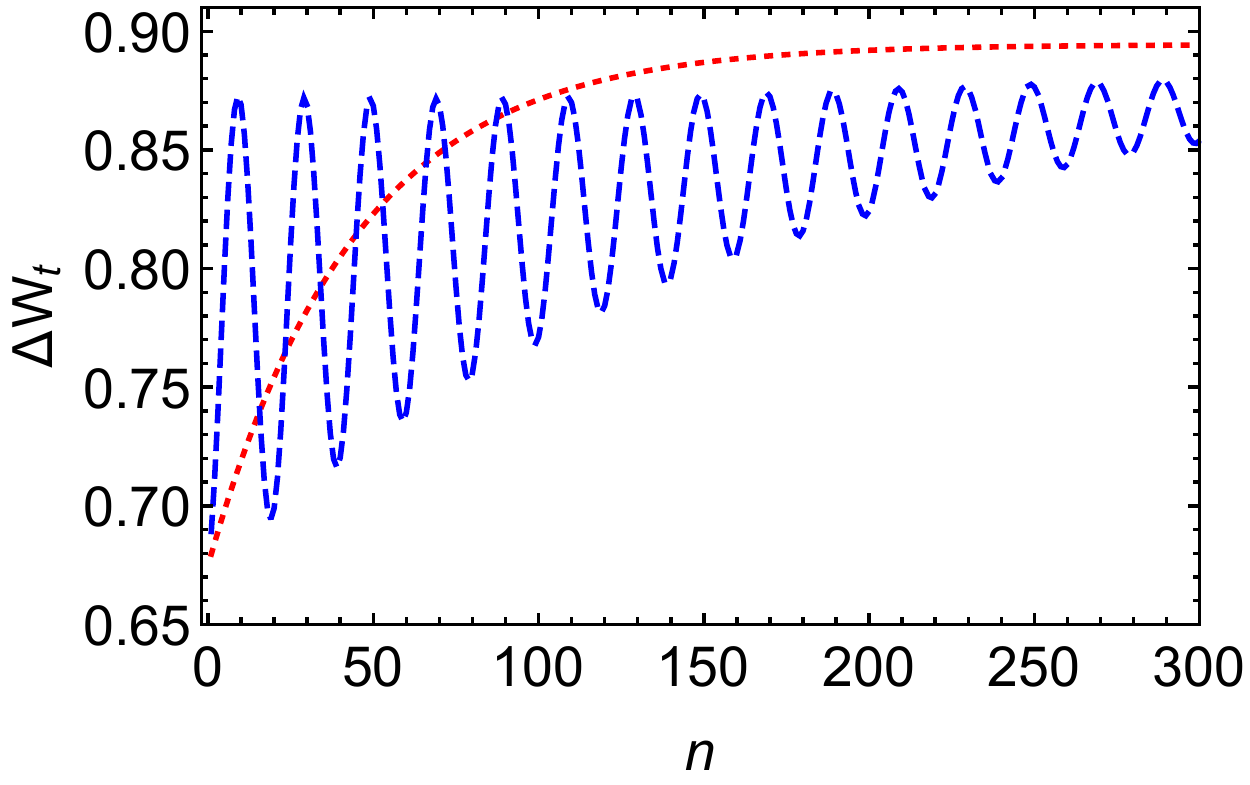}\hspace{0.35cm}\hfill
 \includegraphics[width=160pt]{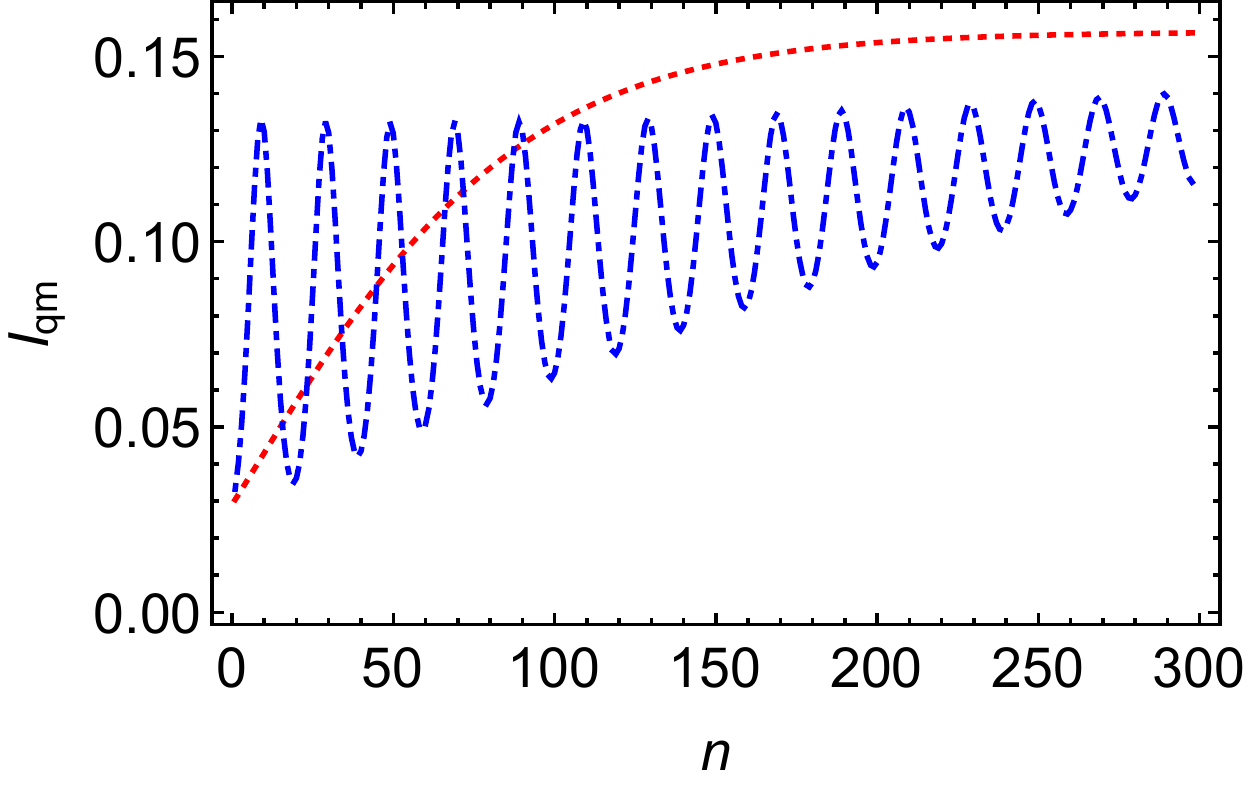}\hspace{0.35cm}\hfill
 \includegraphics[width=160pt]{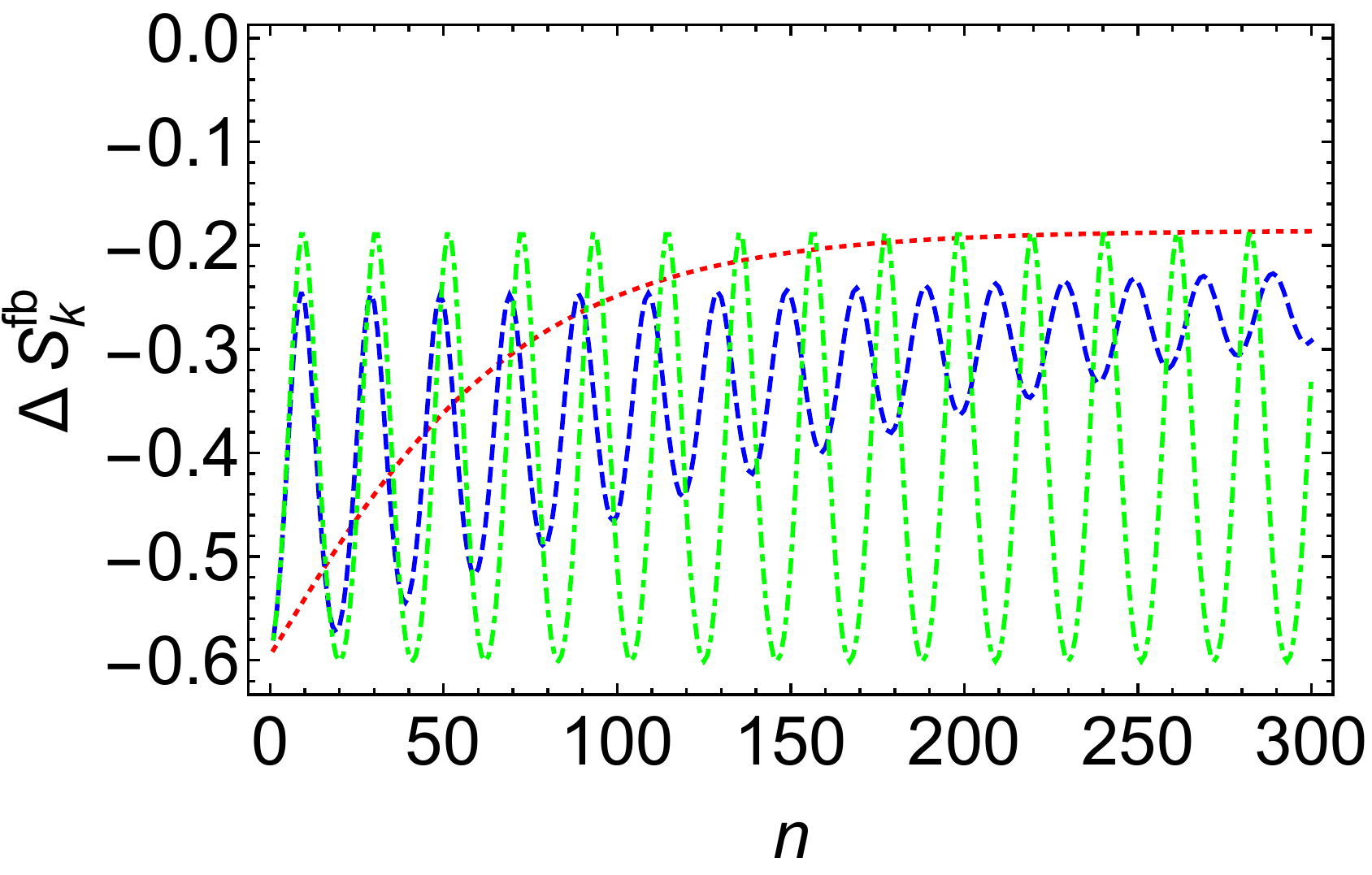}\\
 {\bf (b)} \\
 \includegraphics[width=160pt]{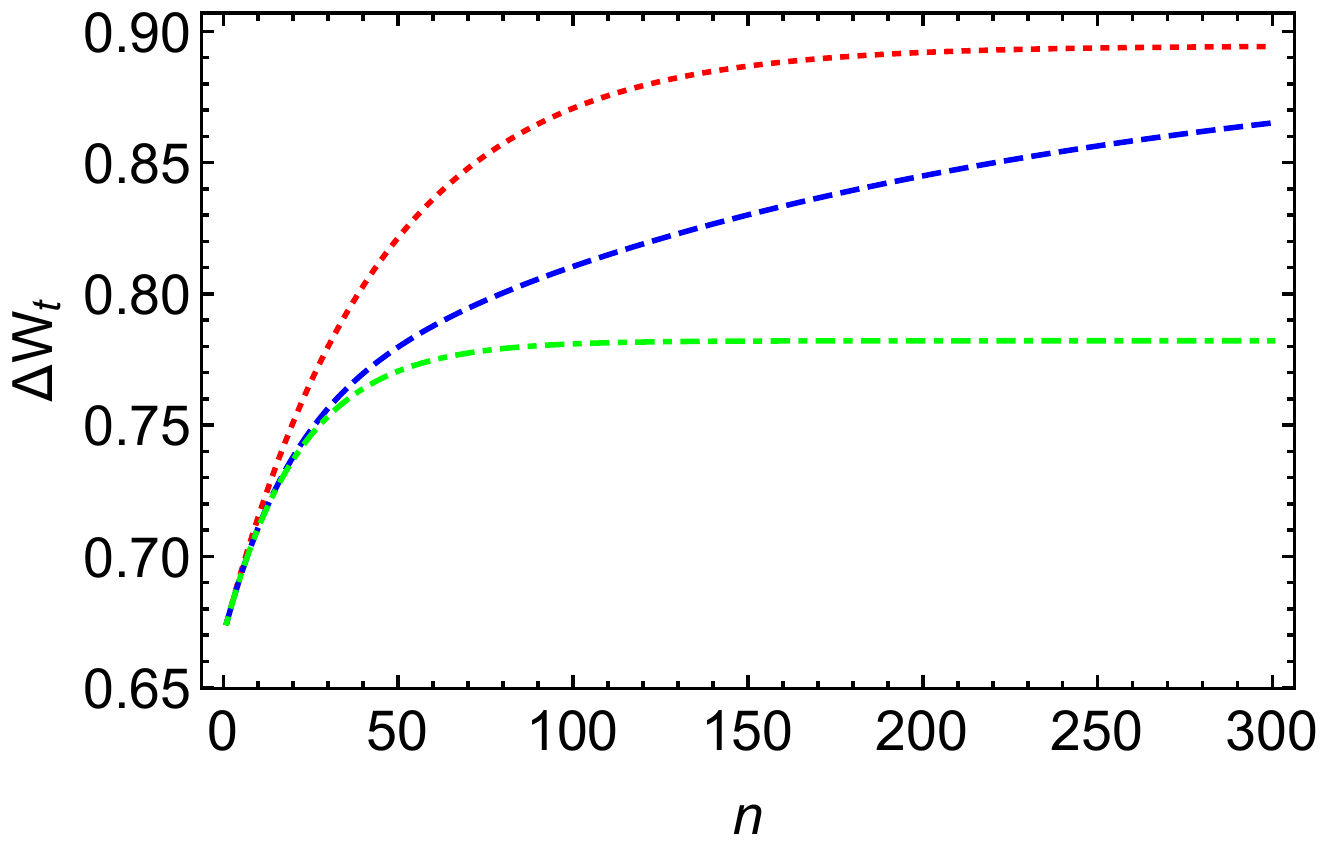}\hspace{0.35cm}\hfill
 \includegraphics[width=160pt]{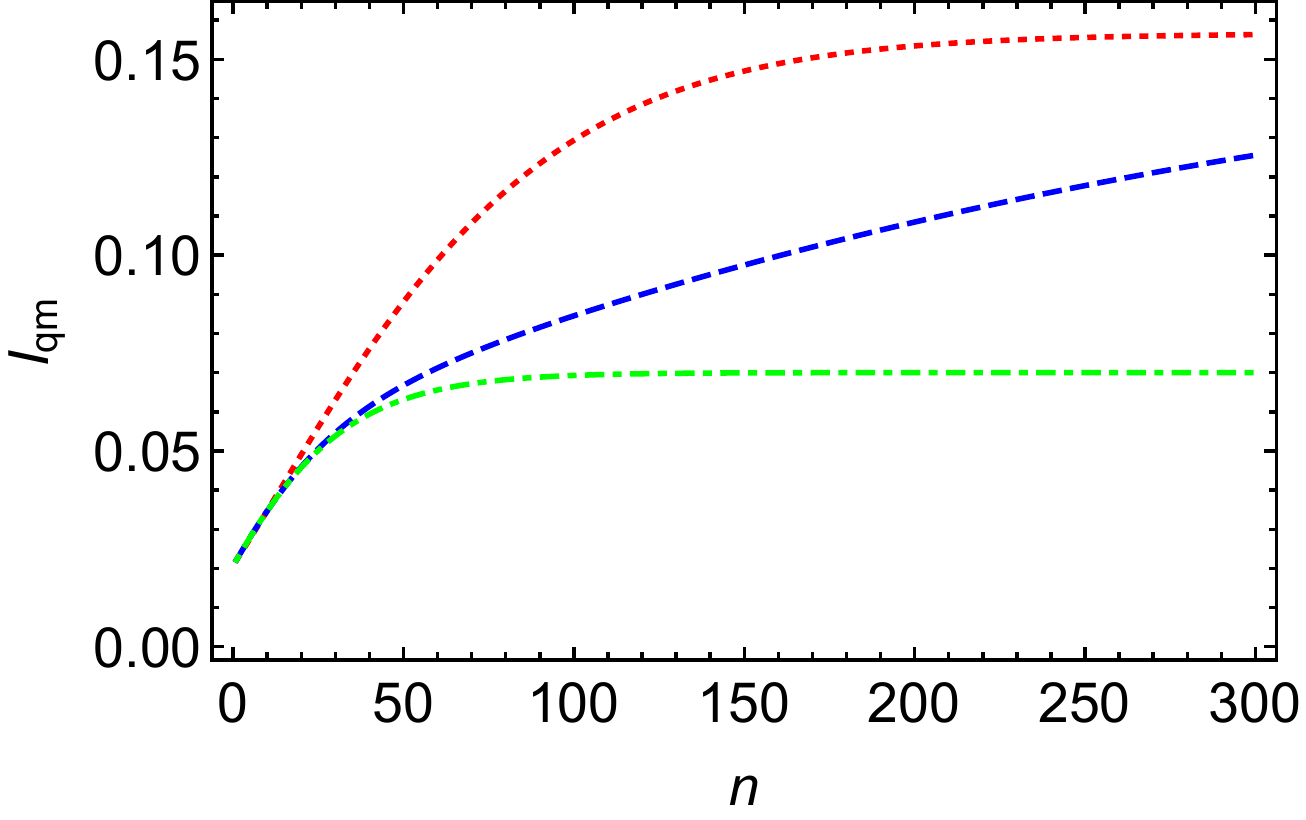}\hspace{0.35cm}\hfill
 \includegraphics[width=160pt]{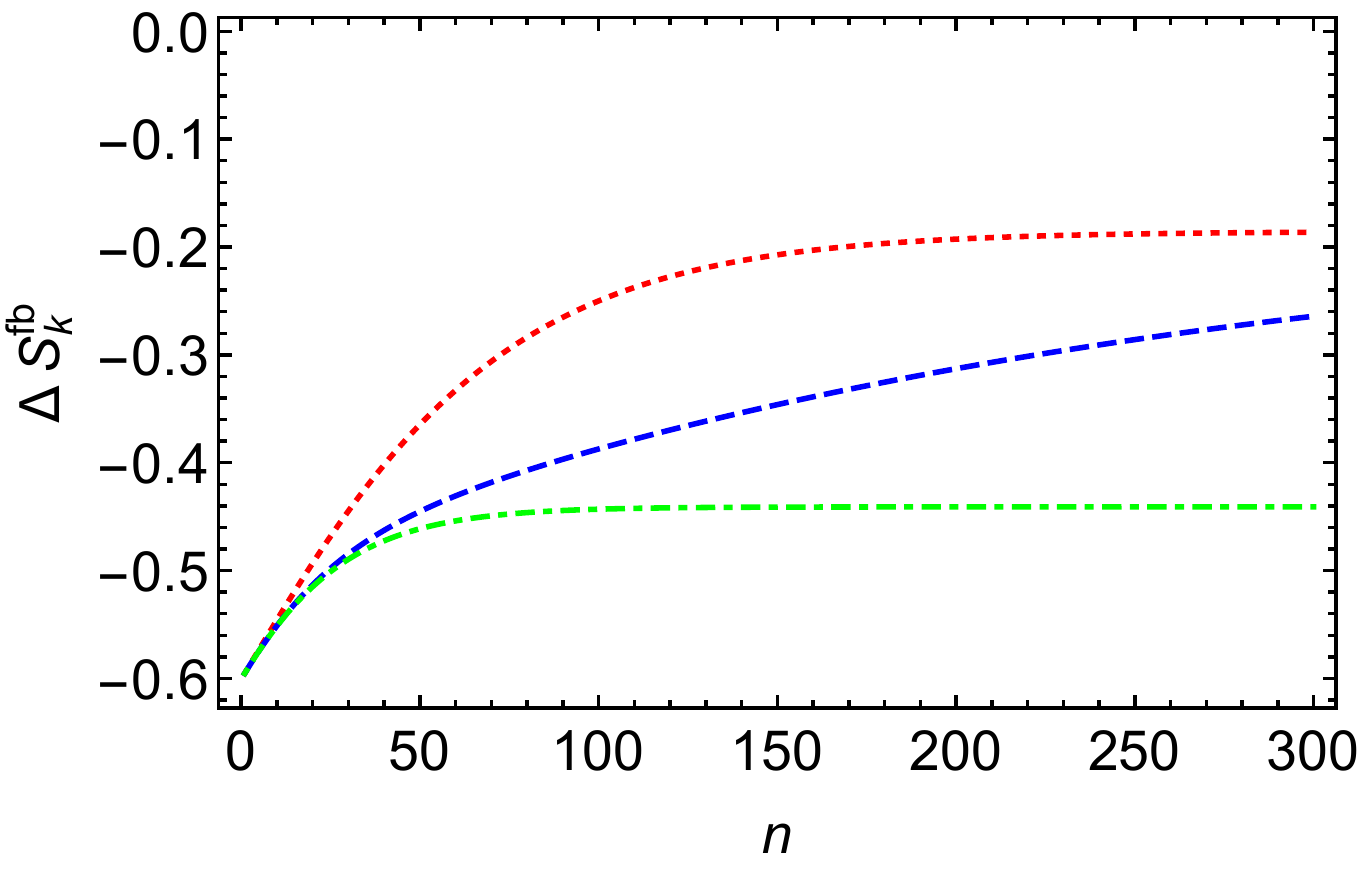}
  \caption{Feedback driven engine performance: The total work extraction $\Delta W_t$,  the quantum mutual information $I_{qm}$ and the entropy change during feedback step $\Delta S_k^{fb}$ as a function of number of   collision $n$ with the environment. The upper panel \textbf{(a)} corresponds to \textit{strategy-1} while the lower panel {\bf (b)} is for \textit{strategy-2}.  The red dotted curve corresponds to the Markovian dynamics, $\tau_e = 0.0$ while the blue dashed curve represent the non-Markovian dynamics, $\tau_e = 10\pi/43$. The green dot-dashed curve represent the full swap non-Markovian dynamics, $\tau_e = \pi/4$. The system-environment interaction time is $\tau = \pi/42$ for weak coupling and  the system and environment frequencies parameters are $\omega_S = 1$ and $\omega_R$ = 3.0 respectively. The system-probe interaction time is  $\tau_m = \pi/14$ and $\beta_S =\beta_R = 0.94$.}
 \label{fig4}
\end{figure*}
\subsection{Feedback-driven engine analysis}\label{section4b}
Let now evaluate the influence of non-Markovianity on  performance of the measurement-based machine described in section \ref{section2} above. We consider a two-level system initially prepared in the state $\rho_S (\sigma_S^z)$ and many identical subenvironment prepared in the state $\rho_R (\sigma_R^z)$.
The measurement ancilla is prepared in the state $\rho_0^M\!=\!\ket{0}\bra{0}$  with the system-measurement apparatus unitary evolution  $U_{SM}$  characterized by the coupling of the form $H_{SM}\!=\!\sigma^x \otimes \sigma^x$. Here, we assumed the $x$-measurement direction but we note that the same optimal value is obtained for $y$-measurement direction considering the initial state of the system. After a feedback operation is performed on the state of the system based on the outcome of the measurement, the thermodynamic quantities, work and quantum mutual information  are numerically calculated, see Fig. \ref{fig4}. Note, in the numerics,  the maximal values of the energy $E[\rho_S^{fb}]$ and entropy $S[\rho_S^{fb}]$ are used and obtained by sampling of the feedback rotation parameters $R(0,\alpha, \theta,\phi)$  from $0 - 2\pi$.

In Fig.~\ref{fig4}, the feedback engine performance, work performed by the engine protocol and the corresponding quantum mutual information associated with the measurement step, as a function of repeated collision are presented for the two different non-Markovian strategies described above. For the Markovian dynamics ($\tau_e = 0$, red dotted curves in Fig. \ref{fig4}(a) and (b)), the work extraction and quantum mutual information increases as the system-environment interactions times grow until it they reach  constant values many collision iteration. For the \textit{strategy-1}, Fig. \ref{fig4}(a), as the system dynamics is prepared to be non-Markovian, an oscillatory  behaviour which vanishes in the long collision  time are observed for both engine performance quantities - work extraction and information gain. The non-Markovian feature is strong at short collision times and can exceed their Markovian counterpart. However, the intermediate system-environment  iteration is marked with suppression of the engine performance due to memory effect. For the non-swap environment-environment interactions (e.g $\tau_e\!=\!10\pi/43$), the work extraction and information gain approach the Markovian values after many number of collisions, see Fig. \ref{fig4}(a). This results from the reduction of information back-flow and the saturation point corresponds to the collision iteration number that the thermodynamic quantities ($\Delta E^u$, $\Delta Q^u$ and $\Delta W^u$) during the preparation step vanishes. In addition, we remark that including the work done on/by the system during the preparation (\textit{step -1}) does not affect our results qualitatively. Moreover, the system entropy change during the feedback $\Delta S^{fb}_k$ exhibit similar behaviour and always negative, see right panel of Fig. \ref{fig4}(a). 

Figure \ref{fig4}(b) shows the work extraction and information gain through measurement resulting from implementation of \textit{strategy-2}. We observe that such non-Markovian dynamics scenario ($\tau_e \ne 0$) gives rise to non oscillatory behaviour  contrary to \textit{strategy-1} and the amount of work extraction and information gain quantities never exceed the Markovian one. This behaviour is akin to the observation in the trace distance Fig.~\ref{fig3}(b), in which the \textit{strategy-2} oscillation are short time leave. Interestingly, for strong environment-environment interaction time $\tau_e\!=\!\pi/4$, the work extraction and information gain saturate to finite value that is lower than the Markovian case, see the green curves in Fig. \ref{fig4}(b). Likewise, the saturation occurs at a vanishing change in the system work done, $\Delta W^u\!=\!0$. For more iterations with fresh environments under weaker interaction environment-environment time $\tau_e\!=\!10\pi/43$, the quantities attain the Markovian values. However, it takes different amount of environment collisions to achieve the Markovian conditions for both strategies.

\section{Conclusion}\label{section5}
We have investigated the interplay between memory effects and performance of a feedback-driven quantum engine. The engine setup consists of  system, reservoir and measurement probe which have modelled as set of two-level systems. We have employed the trace distance as a measure of memory effects (non-Markovianity) to illustrate two strategies of realizing non-Markovian dynamics. We have observed that the memory effect can enhance the performance - \textit{work extraction} and \textit{information gain} of feedback driven engine in a system-environment interaction short time. However, the performance decreases during the intermediate interaction time and approaches the Markovian value at very long time. Besides shedding light on the interplay between non-Markovianity and measurement driven engine, this study suggest more theoretical effort to understand the role of memory on information thermodynamics.

\section*{Acknowledgement}
 We thank Steve Campbell for his helpful comment on the manuscript. The authors acknowledge the support by the Royal Society (Grant Numbers NF160966 and NI160057), the Royal Commission for the Exhibition of 1851, the SFI-DfE Investigator Programme grant (Grant 15/IA/2864), and the H2020 Collaborative Project TEQ (Grant Agreement 766900).


\begin{thebibliography}{100}
\bibitem{lef03} H.S. Leff and A.F. Rex, {\it Maxwell's Demon 2: Entropy, Classical and Quantum Information, Computing} (Princeton University Press, Princeton, NJ, 2003).
\bibitem{szi29} L. Szilard, Z. Phys. 53, 840 (1929).
\bibitem{lan61} R. Landauer, IBM J. Res. Dev. 5, \textbf{183} (1961).
\bibitem{ben82} C.H. Bennett, Int. J. Theor. Phys. 21, \textbf{905} (1982).
\bibitem{par15}  J.M.R. Parrondo, J.M. Horowitz, and T. Sagawa, Nat. Phys. 11, \textbf{131} (2015).
\bibitem{toy10} S. Toyabe, T. Sagawa, M. Ueda, E. Muneyuki, and M. Sano, Nature Phys. \textbf{6}, 988 (2010).
\bibitem{ber12} A. B\'erut, A. Arakelyan, A. Petrosyan, S. Ciliberto, R. Dillenschneider and E. Lutz, Nature \textbf{483} 187 (2012).
\bibitem{kos14} J.V. Koski, V.F. Maisi, J.P. Pekola, and D.V. Averin, Proc. Natl. Acad. Sci. U.S.A. \textbf{111}, 13786 (2014).
\bibitem{kos14a} J.V. Koski, V.F. Maisi, T. Sagawa, and J.P. Pekola, Phys. Rev. Lett. \textbf{113}, 030601 (2014).

\bibitem{bre09} H-P. Breuer, E-M. Laine, and J. Piilo, \PRL \textbf{103}, 210401 (2009).
\bibitem{riv10} A. Rivas, S.F. Huelga and M.B. Plenio, \PRL \textbf{105}, 050403 (2010)
\bibitem{lor13} S.F. Lorenzo, F. Plastina and M. Paternostro, \PRA \textbf{88}, 020102 (2013)
\bibitem{liu11} B-H. Liu, L. Li, Y-F. Huang, C-F. Li, G-C. Guo, E-M. Laine, H-P. Breuer, and J. Piilo, Nature Phys. \textbf{7}, 931
\bibitem{sou13} A.M. Souza, J. Li, D.O. Soares-Pinto, R.S. Sarthour, S. Oliveira, S.F. Huelga, M. Paternostro and F.L. Semi\~ao, arXiv:1308.5761 (2013)
\bibitem{chi12} A. Chiuri, C. Greganti, L. Mazzola, M. Paternostro and P. Mataloni, Sci. Rep. \textbf{2}, 968 (2012).
\bibitem{ree14} D. Reeb and M.M. Wolf, \NJP \textbf{16}, 103011 (2014)
\bibitem{lor15}  S. Lorenzo, R. McCloskey, F. Ciccarello, M. Paternostro and G. M. Palma, Phys. Rev. Lett. {\bf 115}, 120403 (2015).
\bibitem{pez16} M. Pezzutto, M. Paternostro and Y. Omar, \NJP \textbf{18}, 123018 (2016).
\bibitem{ham17} S. Hamedani Raja, M. Borrelli, R. Schmidt, J. P. Pekola, S. Maniscalco, \PRA \text{97}, 032133 (2017).
\bibitem{gel13} D. Gelbwaser-Klimovsky, N. Ezer, R. Alicki, and G. Kurizki, \PRA \textbf{88}, 022112 (2013)
\bibitem{che16} H-B. Chen, P-Y. Chiu and Y-N. Chen, \PRE \textbf{94}, 052101 (2016).
\bibitem{tho18} G. Thomas,  N. Siddharth, S. Banerjee, S. Ghosh, Phys. Rev. E \textbf{97}, 062108 (2018).
\bibitem{jem04} J. Gemmer, M. Michel, and G. Mahler, {\it Quantum Thermodynamics} (Springer, Berlin, 2004).
\bibitem{hub08} G. Huber, F. Schmidt-Kaler, S. Deffner, and E. Lutz, Phys. Rev. Lett. {\bf 101}, 070403 (2008).
\bibitem{cam11} M. Campisi, P. H\"anggi, and P. Talkner, Rev. Mod. Phys. {\bf 83}, 771 (2011).
\bibitem{maz13} L. Mazzola, G. De Chiara, and M. Paternostro, Phys. Rev. Lett. {\bf 110}, 230602 (2013).
\bibitem{bat14} T.B. Batalh\~ao, A.M. Souza, L. Mazzola, R. Auccaise, R.S. Sarthour, I.S. Oliveira, J. Goold, G. De Chiara, M. Paternostro, and R.M. Serra, Phys. Rev. Lett. \textbf{113}, 140601 (2014).
\bibitem{an15} S. An, J.-N. Zhang, M. Um, D. Lv, Y. Lu, J. Zhang, Z.-Q. Yin, H.T. Quan, and K. Kim, Nature Phys. \textbf{11}, 193 (2015).
\bibitem{jar11} C. Jarzynski, Ann. Rev. Cond. Mat. Phys. \textbf{2}, 329 (2011).
\bibitem{sag08} T. Sagawa and M. Ueda, Phys. Rev. Lett. \textbf{100}, 080403 (2008).
\bibitem{sag10} T. Sagawa and M. Ueda, \PRL \textbf{104}, 090602 (2010).
\bibitem{jac09} K. Jacobs,  Phys. Rev. A \textbf{80}, 012322 (2009).
\bibitem{def13} S. Deffner, Phys. Rev. E \textbf{88}, 062128 (2013).
\bibitem{fun13} K. Funo, Y. Watanabe and M. Ueda, \PRE \textbf{88}, 052121 (2013).
\bibitem{hor14} J.M. Horowitz and H. Sandberg, New J. Phys. 16, 125007 (2014)
\bibitem{goo16} J. Goold, M. Huber, A. Riera, L. del Rio, and P. Skrzypzyk, J. Phys. A \textbf{49}, 143001 (2016).
\bibitem{str17} P. Strasberg, G. Schaller, T. Brandes, and M. Esposito, Phys. Rev. X \textbf{7}, 021003 (2017).
\bibitem{cam16} P.A. Camati, J.P.S. Peterson, T.B. Batalh\~ao, K. Micadei, A.M. Souza, R.S. Sarthour, I.S. Oliveira and R.M. Serra,  Phys. Rev. Lett. \textbf{117}, 240502 (2016).
\bibitem{cia17} M. A. Ciampini, L. Mancino, A. Orieux, C. Vigliar, P. Mataloni, M. Paternostro, and M. Barbieri, npj Quantum Information \textbf{3}, 10 (2017).
\bibitem{cot17} N. Cottet,  S. Jezouin,  L. Bretheau,  P. Campagne-Ibarcq, Q. Ficheux, J. Anders, A. Auff\'eves, R. Azouit, P. Rouchon, and B. Huard, Proc.  Natl. Acad. Sci. \textbf{114}, 7561 (2017).
\bibitem{mas17} Y. Masuyama, K. Funo, Y. Murashita, A. Noguchi, S. Kono, Y. Tabuchi, R. Yamazaki, M. Ueda, and Y. Nakamura, arXiv:1709.00548 (2017).
\bibitem{xio18} T.P. Xiong, L.L. Yan, F. Zhou, K. Rehan, D.F. Liang, L. Chen, W.L. Yang, Z.H. Ma, M. Feng, and V. Vedral, \PRL \textbf{120}, 010601 (2018).
\bibitem{nag18} M. Naghiloo, J.J. Alonso, A. Romito, E. Lutz, and K.W. Murch, arXiv:1802.07205 (2018).
\bibitem{boy02} P.O. Boykin, T. Mor, V. Roychowdhury, F. Vatan, and R. Vrijen,  Proc. Natl. Acad. Sci. \textbf{99}, 3388 (2002).
\bibitem{bau05} J. Baugh, O. Moussa, C.A. Ryan, A. Nayak, and R. Laflamme, Nature \textbf{438}, 470 (2005).
\bibitem{liu16} P. Liuzzo-Scorpo, L. A. Correa, R. Schmidt and G. Adesso, Entropy \textbf{18}, 48 (2016).
\bibitem{rod17} N.A. Rodr\'iguez-Briones, E. Mart\'in-Mart\'inez, A. Kempf, and R. Laflamme, Phys. Rev. Lett. \textbf{119}, 050502 (2017).

\bibitem{sca02} V. Scarani, M. Ziman, P. Stelmachovic, N. Gisin, and V. Buzek, \PRL {\bf 88}, 097905 (2002).
\bibitem{zim02} M. Ziman, P. Stelmachovic, V. Buzek, M. Hillery, V. Scarani, N. Gisin, \PRA {\bf 65}, 042105 (2002).
\bibitem{lorenzo} S. Lorenzo, R. McCloskey, F. Ciccarello, M. Paternostro, and G. M. Palma, Phys. Rev. Lett. {\bf 115}, 120403 (2015).
\bibitem{strunz} S. Kretschmer, K. Luoma, and W. T. Strunz, Phys. Rev. A {\bf 94}, 012106 (2016).
\bibitem{ciccarello} F. Ciccarello, Quant. Meas. Quant. Metrol. {\bf 4}, 53 (2017); F. Ciccarello and V. Giovannetti, Phys. Scr. {\bf T153}, 014010 (2013).

\bibitem{sea18} S. Seah and S. Nimmrichter, arXiv:1809.04781 (2018).

\bibitem{mcc14} R. McCloskey and M. Paternostro, Phys. Rev. A {\bf 89}, 052120 (2014).
\bibitem{pezzuttoQST} M. Pezzutto, M. Paternostro, and Y. Omar, Quantum Sci. Technol. {\bf 4}, 025002 (2019).
\bibitem{cam18} S. Campbell, F. Ciccarello, G.M. Palma, and B. Vacchini, Phys. Rev. A {\bf 98}, 012142 (2018). 
\bibitem{cakmakPRA2019} B. Cakmak, S. Campbell, B. Vacchini, \"O.E. M\"ustecaplioglu, and M. Paternostro,  Phys. Rev. A {\bf 99}, 012319 (2019).


\end{thebibliography}
\end{document}